\documentclass[aps,prd,twocolumn,superscriptaddress,preprintnumbers,floatfix,nofootinbib,notitlepage,showkeys,showpacs]{revtex4-1}

\usepackage[utf8]{inputenc}

\usepackage{graphicx}
\usepackage{hyperref}
\usepackage{latexsym}
\usepackage{amsmath}
\usepackage{amssymb}
\usepackage{bbm}
\usepackage{ulem}
\usepackage{pdfsync}
\usepackage{epsfig}
\usepackage{epstopdf}
\usepackage{subfigure}
\usepackage{xcolor}
\usepackage{comment}
\usepackage{slashed}
\usepackage{multirow}
\usepackage{xfrac}

\newcommand{\avg}[1]{\langle #1 \rangle}

\begin{document}

\title{Where are NANOGrav's big black holes?}
\author{Gabriela Sato-Polito}
\email{gsatopolito@ias.edu}
\affiliation{School of Natural Sciences, Institute for Advanced Study, Princeton, NJ 08540, United States}

\author{Matias Zaldarriaga}
\affiliation{School of Natural Sciences, Institute for Advanced Study, Princeton, NJ 08540, United States}

\author{Eliot Quataert}
\affiliation{Department of Astrophysical Sciences, Princeton University, Princeton, NJ 08544, USA}

\begin{abstract}
Multiple pulsar timing array (PTA) collaborations have recently reported the first detection of gravitational waves (GWs) of nanohertz frequencies. The signal is expected to be primarily sourced by inspiralling supermassive black hole binaries (SMBHBs) and these first results are broadly consistent with the expected GW spectrum from such a population. Curiously, the measured amplitude of the GW background in all announced results is a bit larger than theoretical predictions. In this work, we show that the amplitude of the stochastic gravitational wave background (SGWB)  predicted from the present-day abundance of SMBHs derived from local scaling relations is significantly smaller than that measured by the PTAs. We demonstrate that this difference cannot be accounted for through changes in the merger history of SMBHs and that there is an upper limit to the boost to the characteristic strain from multiple merger events, due to the fact that they involve black holes of decreasing masses. If we require the current estimate of the black hole mass density --- equal to the integrated quasar luminosity function through the classic So\l{}tan argument --- to be preserved, then the currently measured PTA result would imply that the typical total mass of SMBHs contributing to the background should be at least $\sim 3 \times 10^{10} M_\odot$, a factor of $\sim 10$ larger than previously predicted. The required space density of such massive black holes corresponds to order 10 $3 \times 10^{10} M_\odot$ SMBHs within the volume accessible by stellar and gas dynamical SMBH measurements. By virtue of the GW signal being dominated by the massive end of the SMBH distribution, PTA measurements offer a unique window into such rare objects and complement existing electromagnetic observations. 
\end{abstract}

\maketitle

Evidence for the first detection of low-frequency ($\sim~1-100$nHz) gravitational waves (GWs) has recently been reported by multiple pulsar timing array (PTA) collaborations. The signature of GWs on the arrival time of pulses is a correlated red noise with a particular angular dependence given by the Helling-Downs \cite{HD} curve. This signature has been detected with varying levels of significance by the European PTA and Indian PTA (\cite{EPTA:2023fyk}, hereafter EPTA+InPTA), by the North American Nanohertz Observatory for Gravitational waves (\cite{NANOGrav:2023gor}, hereafter NANOGrav), Parkes PTA (\cite{Reardon:2023gzh}, hereafter PPTA), and the Chinese Pulsar Timing Array \cite{Xu:2023wog}, that span between $2-4\sigma$. Their results have also been shown to be consist with each other, with discrepancies smaller than $1\sigma$, despite the differences in both observations and modelling choices~\cite{InternationalPulsarTimingArray:2023mzf}.

Although the origin of the signal is unknown, the dominant source of the stochastic gravitational wave background (SGWB) is expected to be a population of inspiralling supermassive black-hole binaries (SMBHBs) (see, e.g., Ref.~\cite{Burke-Spolaor:2018bvk}). Strong observational evidence suggests that most, if not all, galaxies host SMBHs at their centers~\cite{1995ARA&A..33..581K, 1998Natur.395A..14R}. Following the merger of host galaxies, SMBHs form binaries that may also merge~\cite{1980Natur.287..307B}, which would produce the observed PTA signal, albeit through a highly uncertain process. 

The most precise and direct measurements of the SMBH population are derived from observations in our local Universe, where direct kinematic estimates of their masses are possible~\cite{kormendy_ho, mcconnell_ma} for several tens of black holes, along with various properties of the host galaxies. The tight correlations between the SMBH mass and host properties likely suggest a shared evolution and can be extrapolated to the total SMBH population by combining such scaling relations with galaxy catalogs~\cite{2004MNRAS.351..169M, vika2009, shankar2009, shankar2013}. At higher redshifts, inferences of the SMBH population rely on the active black holes powering AGNs. The total mass accumulated in SMBHs from AGN relics can be directly related to the total luminosity they emit through the classic So\l{}tan argument~\cite{soltan1982}, while the cosmic evolution of the SMBH mass function can be theoretically modelled through the continuity equation~\cite{1971ApJ...170..223C, 1992MNRAS.259..725S, 2002MNRAS.335..965Y, 2013MNRAS.428..421S,Tucci:2016tyc}.

The GW spectrum measured by recent PTA analyses is broadly consistent with expectations from an inspiral driven by GW emission \cite{Phinney:2001di}, with the largest deviation in the slope of the spectrum found by NANOGrav. All announced results, however, found amplitudes larger than expected from theoretical predictions~\cite{NANOGrav:2023hfp}. In this work, we use the formalism laid out in Ref.~\cite{Phinney:2001di} to connect the binaries contributing to the SGWB to the present-day population of remnant (single) black holes in order to assess the significance of this discrepancy. This approach yields a simple but robust model for the SGWB that sheds light on the uncertainties associated with the merger history of SMBHBs, which enables a clear comparison between PTAs and other observations of SMBHs.

We show that the difference between predicted and measured values of the SGWB cannot be accounted for by an increased black hole merger rate for a SMBH mass function consistent with local observations. We demonstrate that there is a mathematical upper limit to the boost of the characteristic strain from multiple mergers, which is only a factor of $\sim 1.64$ due to the fact that they involve black holes of decreasing masses. Since the predicted amplitude of the SGWB falls short of the measured value even in the most ideal scenario, we explore modifications to the mass function prediction that could lead to this difference.

If the current estimate of the mass density in black holes $\rho_{\rm BH}$, consistent with the So\l{}tan argument, is to be preserved, we argue that all modifications to the SMBH mass function are equivalent to a change in the peak mass contributing to the SGWB. The PTA measurements would therefore generally imply that the SGWB is dominated by SMBHs roughly 10 times larger than previously estimated, of around $M^{\rm GW}_{\rm peak} \sim 3 \times 10^{10} M_{\odot}$. We begin by investigating whether a higher intrinsic scatter $\epsilon_0$ in the $M-\sigma$ relation or a higher value for the turnover $\sigma_*$ in the velocity dispersion function might be responsible for this misestimation, but find that the required values are in significant disagreement with current observations. 

Another reason why the peak BH mass contributing to the SGWB may be larger is if the $M-\sigma$ relation becomes steeper for high velocity dispersion galaxies. Indeed, although there are too few objects to derive definitive conclusions, the $M-\sigma$ does appear to steepen for $\sigma\gtrsim 275$km s$^{-1}$\cite{kormendy_ho, mcconnell_ma}, a consequence of the galaxy population ``saturating" at high $\sigma$ \cite{2013ApJ...769L...5K, Thomas:2013bga, 2012MNRAS.424..224H}. The currently favored picture is that these massive galaxies that host the most massive SMBHs have cores that are formed in the aftermath of gas-poor major mergers by SMBHBs, which scour the stars at the center of the merged galaxy~\cite{1991Natur.354..212E, 1997AJ....114.1771F, 2001ApJ...563...34M,2019ApJ...887...10S}. This scenario suggests that the high-mass end of the SMBH mass function would be underpredicted by the $M-\sigma$ relation, but would be more accurately estimated from the luminosity or bulge mass relations. We show, however, that the amplitude of the SGWB predicted from the relation between black hole mass and bulge mass also leads to a smaller value than observed, thus suggesting that this effect is not sufficient to account for the discrepancy. If a larger characteristic strain is to be explained by a steeper $M-\sigma$ relation, it must therefore necessarily imply either a larger abundance of high stellar mass galaxies or a corresponding change to the $M-M_{\rm bulge}$ relation. Finally, we also consider a more general modification to the SMBH mass function, in which we simply boost the mass of the black hole associated with galaxies of a given velocity dispersion. We show what the required black hole mass is for each value of $\sigma$, suggesting that the difference may be explained by a small number of extremely massive black holes or a large number of lighter ones.

This paper is organized as follows. In Sec.~\ref{sec:scaling} we discuss the SMBH population inferred from local scaling relations coupled to galaxy catalogs, as well as the information offered by the quasar luminosity function about this population. We discuss the model for the characteristic strain in Sec.~\ref{sec:PTA} and how it depends on the present-day SMBH population presented in the previous section. In Sec.~\ref{sec:discussion} we show the discrepancy between the two sets of observations and discuss potential sources of this difference, and conclude in Sec.~\ref{sec:conclusion}.

\section{Black hole population from scaling relations and AGNs}\label{sec:scaling}
For nearby sources, black hole masses can be directly estimated from gas or stellar kinematics, along with host galaxy properties such as velocity dispersion, bulge stellar mass, and luminosities. The current sample contains several tens of galaxies \cite{kormendy_ho, mcconnell_ma} and establishes tight relations between supermassive black holes and their hosts, which are given by power laws between host properties ($X$) and the black hole mass ($M_{\rm BH}$)
\begin{equation}
    \log_{10}M_{\rm BH} = a_{\bullet} + b_{\bullet}\log_{10} X,
    \label{eq:M-X}
\end{equation}
with a log-normal intrinsic scatter of $\epsilon_0$. In order to infer the total SMBH population, these scaling relations must then be combined with galaxy catalogs. 

Our fiducial model relies on the galaxy stellar velocity dispersion ($\sigma$) as a proxy for black hole mass. We combine the $M-\sigma$ scaling relation above (with $X=\sigma/200$km s$^{-1}$) and combine with the galaxy velocity dispersion function (VDF), typically parametrized as
\begin{equation}
    \phi(\sigma) d\sigma = \phi_{*} \left(\frac{\sigma}{\sigma_*}\right)^{\alpha} \frac{e^{-(\sigma/\sigma_*)^{\beta}}}{\Gamma(\alpha/\beta)} \beta \frac{d\sigma}{\sigma},
    \label{eq:sigma_function}
\end{equation}
which corresponds to a generalization of the Schechter function. While many host galaxy properties have been shown to correlate with the mass of the central SMBH, recent work suggests that velocity dispersion offers a more reliable, and perhaps more fundamental, prediction of black hole mass \cite{kormendy_ho, 2016ApJ...831..134V, Marsden:2020yyd}. However, this may not necessarily hold for the most massive SMBHs --- the velocity dispersion of galaxies appears to saturate above $\sigma \gtrsim 275$km s$^{-1}$, which might lead to an underprediction of the massive end of the SMBH mass function relative to bulge mass ($M_{\rm bulge}$) or luminosity scaling relations \cite{2007ApJ...662..808L, 2019ApJ...887...10S}. Hence, while we focus on the velocity dispersion in this discussion, we also estimate the SMBH mass function using the bulge mass scaling relation. 

In the absence of scatter in the $M-\sigma$ relation, the SMBH mass function can be computed by directly combining Eqs.~\ref{eq:M-X} and \ref{eq:sigma_function}. Substituting velocity dispersion for the black hole mass leads to
\begin{equation}
    \phi(M_{\rm BH}) dM_{\rm BH} = \phi_* \left(\frac{M_{\rm BH}}{M_*}\right)^{\alpha/b_{\bullet}} \frac{e^{-(\frac{M_{\rm BH}}{M_*})^{\beta/b_{\bullet}}}}{\Gamma(\alpha/\beta)} \frac{\beta}{b_{\bullet}} \frac{dM_{\rm BH}}{M_{\rm BH}},
    \label{eq:BHMF}
\end{equation}
where we have defined $\log_{10}M_{*} = a_{\bullet} + b_{\bullet}\log_{10}\sigma_*/200$km~s$^{-1}$. In the presence of scatter, we now have that the mass function is given by a convolution between the velocity dispersion function and the probability distribution function $p(\log_{10}M_{\rm BH}|\log_{10}\sigma)$
\begin{equation}
    \phi(M_{\rm BH}) = \int d\sigma \frac{p(\log_{10}M_{\rm BH}|\log_{10}\sigma)}{M_{\rm BH} \log(10)} \phi(\sigma),
    \label{eq:MF_scatter}
\end{equation}
where we assume $p$ to be log-normal
\begin{equation}
\begin{split}
    p(\log_{10}&M_{\rm BH}|\log_{10}\sigma) = \frac{1}{\sqrt{2\pi} \epsilon_0} \\ &\times \exp\left[ -\frac{1}{2}\left(\frac{\log_{10}M_{\rm BH} - a_{\bullet}-b_{\bullet}\log_{10}\sigma}{\epsilon_0}\right)^2 \right].
\end{split}
\end{equation}

\begin{figure}
    \centering
    \includegraphics[width=0.48\textwidth]{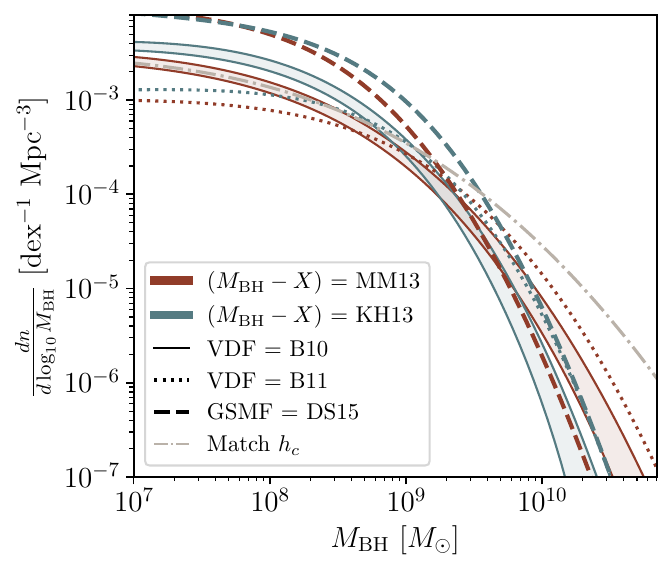}
    \caption{SMBH mass function computed from a combination of VDFs and GSMF with local scaling relations reported in the literature. The VDFs correspond to Refs.~\cite{2010MNRAS.404.2087B} and \cite{2011ApJ...737L..31B}(B10 and B11, respectively), the GSMF is given in \cite{2015MNRAS.454.4027D} (DS15), while the adopted scaling relations are presented in Refs.~\cite{kormendy_ho} and \cite{mcconnell_ma} (KH13 and MM13, respectively). For the sake of visual clarity, we have omitted the error bands of most SMBH mass functions in the plot, and show only for the B10 VDF result since it corresponds to our fiducial choice. The grey dashdotted line shows an example of a SMBH mass function that matches the characteristic strain observed by PTAs, computed by increasing the intrinsic scatter to $\epsilon_0=0.62$.}
    \label{fig:BHMF}
\end{figure}

We predict the SMBH mass function from various combinations of velocity dispersion functions and $M-\sigma$ relations, which we describe next. Ref.~\cite{2010MNRAS.404.2087B} (B10) uses spectroscopic data from SDSS of $\sim 250000$ galaxies across $4681$deg$^2$ of the sky and redshifts $z\lesssim 0.3$. We choose the VDF fit to all galaxies and including only $\sigma>125$km s$^{-1}$, since the low-velocity dispersion end is subject to greater uncertainties and systematic errors (e.g. sample completeness), but are not expected to host SMBHs that significantly contribute to the SGWB. Ref.~\cite{2011ApJ...737L..31B} (B11) uses photometric data from the UKIDSS Ultra-Deep Survey (UDS) and the NEWFIRM Medium Band Survey of the COSMOS field (NMBS COSMOS), covering 0.77deg$^2$ and 0.21deg$^2$, respectively. Stellar velocity dispersions are estimated from the radius, Sérsic index, and stellar mass using the virial theorem. We use the results measured between redshift $0.3\leq z \leq 0.6$ and fit the generalized Schechter function given in Eq.~\ref{eq:sigma_function}. Finally, Ref.~\cite{2022ApJ...939...90T} (T20) reports spectroscopic measurements of the VDF from the Large Early Galaxy Astrophysics Census (LEGA-C) survey, which targets massive galaxies in the COSMOS field, from which we use the measurements between $0.6<z \leq 0.7$ to fit Eq.~\ref{eq:sigma_function}. Despite the differences in redshift, Refs.~\cite{2011ApJ...737L..31B} and \cite{2022ApJ...939...90T}, which explore broader redshift ranges, do not find evidence for substantial evolution in the VDF, and we therefore compare the three different measurements as representatives of the present-day VDF. We use two different fits to the $M-\sigma$ relation from Refs.~\cite{kormendy_ho} (KH13) and ~\cite{mcconnell_ma} (MM13, fit to all galaxies).

We also estimate the SMBH mass function using the bulge mass scaling relation. We use the galaxy stellar mass function (GSMF) from Ref.~\cite{2015MNRAS.454.4027D} (DS15), which was measured from a complete sample of half a million galaxies from the SDSS survey. A significant source of systematic error in measurements of the stellar mass function arises in the determination of the total flux in a given bandpass and the impact of sky background subtraction. Ref.~\cite{2015MNRAS.454.4027D} focuses on estimating flux corrections by stacking samples of similar galaxies selected by various properties, while accounting for a variety of potential sources of systematic biases. The measured stellar mass function lies roughly in between previously published results~\cite{2009MNRAS.398.2177L, Bernardi:2013mqa}. We combine this GSMF with the scaling relations from MM13 estimated from bulges with dynamical masses, and from KH13, where we assume a bulge fraction of 1.

In Fig.~\ref{fig:BHMF}, we show multiple estimates of the SMBH mass function using a variety of different types of observations. All choices lead to similar black-hole abundances for the relevant mass range probed by the SGWB (which, as we show in the next section, is around $\sim 3\times 10^{9} M_{\odot}$). For the estimates based on velocity dispersion, this can be attributed to the fact that direct mass measurements for such objects are relatively robust and correspond to galaxies with $\sigma \sim 300$km s$^{-1}$. Low velocity dispersion galaxies $\sigma \lesssim 150$km s$^{-1}$ are impacted by sample completeness, while galaxies with high velocity dispersions $\sigma \gtrsim 400$km s$^{-1}$ are rare and few objects can be found in observed catalogs. The typical galaxies contributing to the SGWB are not impacted by either effects or by morphological selection criteria, being dominated by early-type galaxies.

We choose as our fiducial model the combination of the VDF from B10 and the $M-\sigma$ relation from MM13. In summary, $X=\sigma/200$km s$^{-1}$ in Eq.~\ref{eq:M-X} and we have chosen fiducial parameters: $\phi_*=(2.61 \pm 0.16)\times 10^{-2}$Mpc$^{-3}$, $\sigma_*=159.6\pm1.5$ km s$^{-1}$, $\alpha=0.41\pm 0.02$, and $\beta=2.59\pm 0.04$ for the VDF, and $a_{\bullet}= 8.32\pm 0.05$, $b_{\bullet}= 5.64\pm 0.32$, and $\epsilon_0=0.38$, for the $M-\sigma$ relation. 

Beyond our local Universe, only active SMBHs powering active galactic nuclei (AGN) can be observed. Assuming that the present-day population of SMBHs are relics of AGNs, the total mass density accumulated in black holes can be related to the energy density emitted by AGNs through the So\l{}tan argument \cite{soltan1982}. If all black holes accrete with a constant radiative efficiency $\epsilon_r$, the quasar bolometric luminosity ($L_{\rm bol}$) can be related to the SMBH accretion rate
\begin{equation}
    L_{\rm bol} = \epsilon_r \dot{M}c^2,
\end{equation}
and the integrated quasar luminosity function therefore corresponds to the integrated mass density in SMBHs 
\begin{align}
    \rho_{\rm BH} &= \int dM_{\rm BH} \phi(M_{\rm BH}) M_{\rm BH} \label{eq:rhoBH} \\ &=\frac{1-\epsilon_r}{\epsilon_r c^2} \int_0^{\infty} dz \frac{dt}{dz} \int d\log L_{\rm bol} \phi(L_{\rm bol}, z) L_{\rm bol},
\end{align}
with a relatively consistent value found throughout the literature \cite{2002MNRAS.335..965Y, 2004MNRAS.351..169M, shankar2009, 2007ApJ...654..731H}. For instance, Ref.~\cite{2007ApJ...654..731H} found a value of
\begin{equation}
    \rho_{\rm BH} = 4.81^{+1.24}_{-0.99}\left(\frac{0.1}{\epsilon_r}\right)\times 10^5 M_{\odot} \text{Mpc}^{-3},
\end{equation}
where the radiative efficiency is typically assumed to be a free parameter, fit to explain the relic black hole population. 

\section{Black hole population from PTAs}\label{sec:PTA}
The characteristic strain of the gravitational-wave background produced by inspiralling SMBHBs can be written as the integral of the emission from binaries belonging to each mass and redshift interval, summed over the entire population of sources. That is,
\begin{equation}
    h^2_c(f) = \frac{4G}{\pi c^2 f^2}\int d\mathcal{M} \int \frac{dz}{1+z} \frac{dn}{dzd\mathcal{M}} \left. \left(\frac{d E_{\rm gw}}{d\log f_r}\right)\right|_{f_r=(1+z)f}
\label{eq:h2c}
\end{equation}
where $f_r$ and $f$ are the emitted and observed GW frequencies, and $dn/dzd\mathcal{M}$ is the number of black hole binaries from redshifts between $[z, z+dz]$ and chirp masses $[\mathcal{M}, \mathcal{M}+ d\mathcal{M}]$. For a circular orbit in the Newtonian regime, the energy spectrum depends only on the chirp mass of the binary, and is given by
\begin{equation}
    \frac{dE_{\rm gw}}{d\log f_r} = \frac{1}{3G} (G\mathcal{M})^{5/3} (\pi f_r)^{2/3}.
\end{equation}

Switching from chirp mass to total mass $M_{\rm BH}$, and defining
\begin{equation}
    \epsilon({M_{\rm BH}}) \equiv  \frac{\eta}{3} \left(\frac{v}{c}\right)^2 = \frac{\eta}{3} \frac{(G M_{\rm BH} \pi f)^{2/3}}{c^2},
\end{equation}
where $\eta=q/(1+q)^2$ is the symmetric mass ratio and $q=m_1/m_2$. Using this definition, the characteristic strain can be rewritten as
\begin{equation}
\begin{split}
    h^2_c(f) =& \frac{4 G}{\pi f^2} \int dM_{\rm BH} \int \frac{dz}{(1+z)^{1/3}} \int dq \frac{dn}{dz dM_{\rm BH} dq} \\ &\times M_{\rm BH} \epsilon({M_{\rm BH}}, q)
\end{split}
\end{equation}
We can interpret $\epsilon$ as an average efficiency of GW emission.

\begin{figure}
    \centering
    \includegraphics[width=0.43\textwidth]{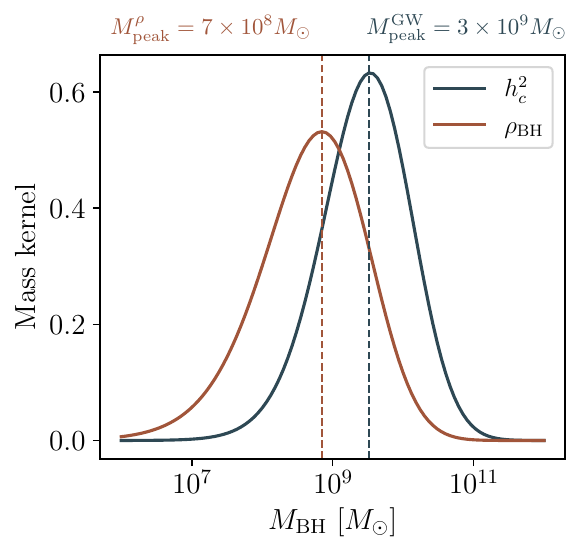}
    \caption{Contribution to the mass density in black holes (red) and to the SGWB (dark blue) per logarithmic mass bin, assuming the fiducial set of parameters (which includes scatter in the $M-\sigma$ relation). The peak of the each kernel is marked by the vertical dashed lines and the exact values are quoted above.}
    \label{fig:kernel}
\end{figure}

A crucial step, pointed out in Ref.~\cite{Phinney:2001di}, is that the remnant black hole population --- resultant from the merger history described by $\frac{dn}{dz dM_{\rm BH} dq}$ --- must correspond to the present-day single SMBH population. Assuming all black holes have undergone a single merger event in their lifetimes, the mass function of mergers integrated over all redshifts and mass ratios must correspond to the single BH mass function ($\phi$) today 
\begin{equation}
    \phi(M_{\rm BH}) = \frac{dn}{dM_{\rm BH}}= \int dz \int dq \frac{dn}{dz dM_{\rm BH} dq}.
\end{equation}
We will demonstrate in Sec.~\ref{sec:merger_history} that the assumption of a single merger event provides a reasonable estimate of the SGWB, since repeated mergers do not significantly increase GW emission. This is due to the fact that each earlier merger must occur between black holes of decreasing masses and therefore contribute progressively less to the characteristic strain.

Assuming that the mass and redshift dependence are separable, such that $\frac{dn}{dz d M_{\rm BH} dq} = p_z(z) p_q(q) \frac{dn}{dM_{\rm BH}}$, we can write the characteristic strain of the SGWB in terms of the local SMBH mass function
\begin{equation}
    h^2_c(f) =\frac{4 G}{\pi f^2} \avg{(1+z)^{-1/3}}\int dM_{\rm BH} \phi(M_{\rm BH}) M_{\rm BH} \avg{\epsilon}({M_{\rm BH}}),
    \label{eq:h2c_simplified}
\end{equation}
where we have defined the redshift and mass ratio averages
\begin{equation}
    \avg{(1+z)^{-1/3}} = \int dz \frac{p_z(z)}{(1+z)^{1/3}}
\end{equation}
and
\begin{equation}
    \avg{\epsilon}({M_{\rm BH}})= \int dq\ p_q(q) \epsilon({M_{\rm BH}}, q).
\end{equation}
We choose the following parametrizations for the redshift and mass-ratio distribution functions
\begin{equation}
    p_z(z) = z^\gamma e^{-z/z_*}, \quad \text{and} \quad p_q(q) = q^{\delta},
    \label{eq:pz_pq}
\end{equation}
with fiducial values of $\gamma=0.5$, $z_*=0.3$, and $\delta=-1$, an assumed minimum value of $q_{\rm min}=0.1$, and where both functions are then normalized to unity. These values are chosen to roughly reproduce the simulation-based results of Ref.~\cite{Kelley:2016gse}, such that the median redshift of the SGWB signal is sourced at $z\sim 0.3$ and over $90\%$ of the contribution comes from $z\lesssim 1$. Notice that the redshift integral is very insensitive to variations in the probability distribution of mergers. Our fiducial set of parameters leads to $\avg{(1+z)^{-1/3}} =0.9$, while its upper limit is 1 and, for instance, a much broader redshift distribution with $z_*=2$ results in $\avg{(1+z)^{-1/3}} = 0.68$. The maximum efficiency of $\eta = 0.25$ corresponds to equal mass ratio mergers, while our fiducial log-uniform distribution leads to $\avg{\eta} = 0.18$.

Using the black hole mass function predicted from Eq.~\ref{eq:MF_scatter} and our fiducial set of parameters, we show in Fig.~\ref{fig:kernel} the contribution to the BH mass density and the characteristic strain per logarithmic mass. This illustrates that there is a characteristic mass that primarily dictates the observed signals and that the SGWB receives the majority of its contribution from significantly higher masses than the mass density, since it is weighted by a factor of $M_{\rm BH}^{5/3}$. 

\begin{figure}[t]
    \centering
    \includegraphics[width=0.5\textwidth]{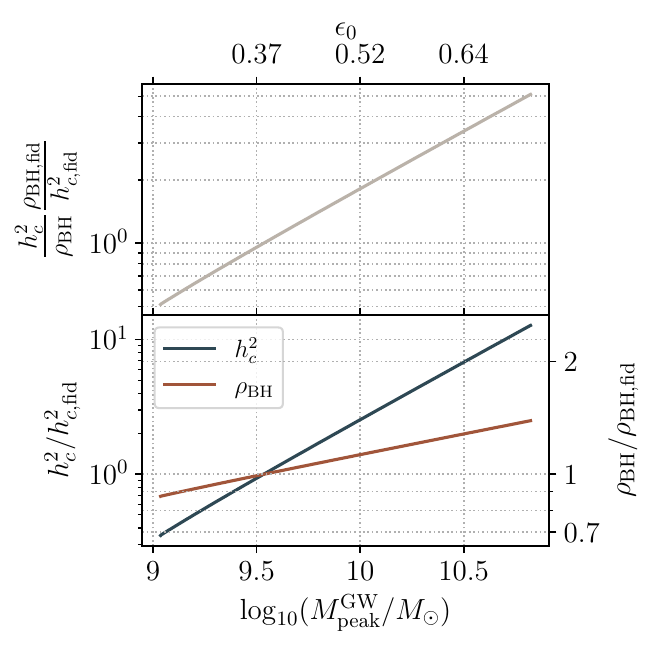}
    \caption{Characteristic strain, black hole mass density, and their ratios as a function of the SMBH mass that most contributes to the SGWB ($M^{\rm GW}_{\rm peak}$), and the corresponding intrinsic scatter $\epsilon_0$ in the upper axis. In the bottom panel, the value of the ratio between characteristic strain and the fiducial estimate are shown in the left axis, while the black hole mass density is shown on the right.}
    \label{fig:hc_rho_ratio}
\end{figure}

The black hole mass density and characteristic strain can now be computed using Eqs.~\ref{eq:rhoBH} and \ref{eq:h2c_simplified} by directly integrating over the BH mass function, 
\begin{align}
    \rho_{{\rm BH}, 0} =& \phi_* M_{*}\frac{\Gamma\left(\sfrac{(\alpha+b_{\bullet})}{\beta}\right)}{\Gamma\left(\sfrac{\alpha}{\beta}\right)}, \label{eq:rhoBH0} \\
    h^2_{c,0}(f)=& \frac{4G \avg{(1+z)^{-1/3}}}{\pi c^2 f^2}  \rho_{{\rm BH}, 0} \epsilon(M_*) \frac{\Gamma\left(\sfrac{(\alpha+5b_{\bullet}/3)}{\beta}\right)}{\Gamma\left(\sfrac{(\alpha+b_{\bullet})}{\beta}\right)},\label{eq:h2c0}
\end{align}
where the gamma function $\Gamma$ results from the mass integral and the assumed functional form of the velocity dispersion function, and the subscript $0$ denotes the value without scatter in the $M-\sigma$ relation. 

By combining Eqs.~\ref{eq:rhoBH} and \ref{eq:h2c_simplified} with \ref{eq:MF_scatter} we can show that scatter boosts the mass density by a factor of $\rho_{\rm BH} = \rho_{\rm BH, 0} e^{\epsilon_0^2 \log^2(10)/2}$ and the characteristic strain by $h^2_c=h^2_{c,0} e^{25/18 \epsilon_0^2 \log^2(10)}$. We find the following values for the black hole mass density and characteristic strain
\begin{equation}
\begin{split}
    \rho_{\rm BH} =& 4.5\times 10^{5}M_{\odot}\text{Mpc}^{-3} \left( \frac{M_{*}}{5.8\times 10^7 M_{\odot}} \right) \\ & \times\left( \frac{1.5}{e^{\frac{1}{2} \epsilon_0^2 \log^2(10)}} \right) \\
    h^2_c(f)=& 1.2\times 10^{-30} \left(\frac{f}{\text{yr}^{-1}}\right)^{-4/3} \left(\frac{M_{*}}{5.8\times 10^7 M_{\odot}}\right)^{2/3} \\ &\times \left(\frac{\rho_{\rm BH}}{4.5\times 10^{5}M_{\odot}\text{Mpc}^{-3}}\right) \left( \frac{2}{e^{\frac{8}{9} \epsilon_0^2 \log^2(10)}} \right).
\end{split}
\end{equation}
Notice that the mass scale $M_*$ corresponds to the turnover in the velocity dispersion function, but it is not the mass that most contributes to the SGWB or the mass density. These were shown in Fig.\ref{fig:kernel} and can be derived analytically in absence of scatter from Eqs.~\ref{eq:BHMF}, \ref{eq:rhoBH0}, and \ref{eq:h2c0}, by computing the peak of the mass integral kernel. The mass that most contributes to the BH mass density and SGWB, per logarithmic mass bin, are given by
\begin{equation}
\begin{split}
    M^{\rho}_{\rm peak} =& M_* \left[ \frac{(\alpha + b_{\bullet})}{\beta} \right]^{b_{\bullet}/\beta} \\
    M^{\rm GW}_{\rm peak} =& M_* \left[ \frac{(\alpha + \sfrac{5 b_{\bullet}}{3} )}{\beta} \right]^{b_{\bullet}/\beta},
\end{split}
\end{equation}
which then scales as $\log M_{\rm peak} \propto \epsilon^2_0$ in the presence of scatter.

One interesting quantity to consider is the ratio between the characteristic strain squared and the black hole mass density. Although the mass density can only be inferred from the quasar luminosity function up to a constant factor given by the radiative efficiency, the ratio $h^2_c/ \rho_{\rm BH}$ cancels the amplitude of the mass function and therefore any dependence on $\epsilon_r$. If we therefore consider a black hole mass function that leads to a $\rho_{\rm BH}$ consistent with the So\l{}tan argument for some value of $\epsilon_r$, and a value of $h_c^2$, then this ratio can only be altered through changes to the mass function that lead to a different $M_{\rm peak}^{\rm GW}$. Increasing the characteristic strain while maintaining the current estimate of $\rho_{\rm BH}$ therefore implies modifications to the high-mass tail of the SMBH mass function, that can be generally described as a change to the peak mass contributing to the SGWB. The ratio as a function of the peak mass is shown on the top panel of Fig.~\ref{fig:hc_rho_ratio}, while $h^2_c$ and $\rho_{\rm BH}$ are shown on the bottom. 

\begin{figure}
    \centering
    \includegraphics[width=0.45\textwidth]{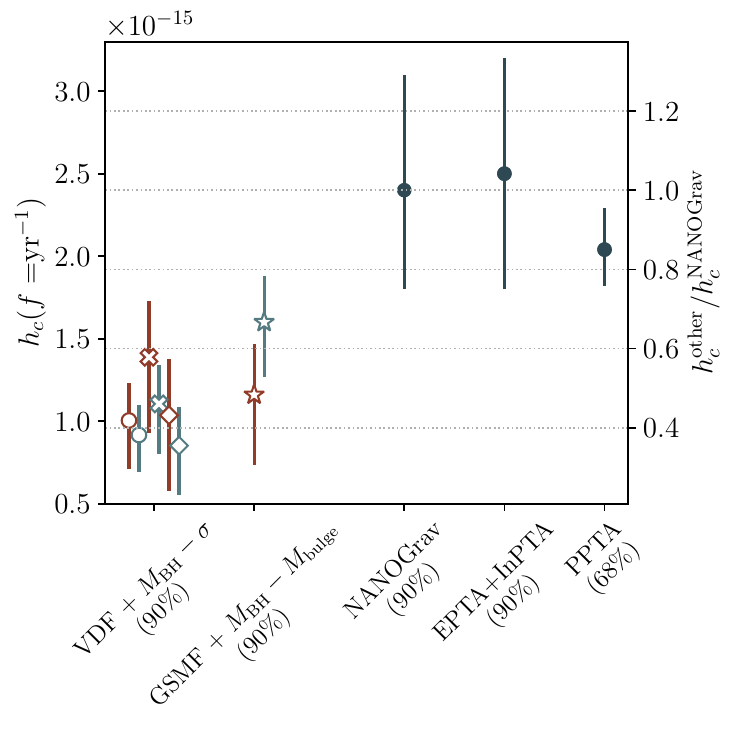}
    \caption{Comparison between the characteristic strain predicted using galaxy scaling relations and galaxy catalogs (open symbols), and the values reported by PTA collaborations (filled dots). The left vertical axis shows the characteristic strain value at the reference frequency of $f=$yr$^{-1}$ for a fixed power law of $\gamma=13/3$, while the right vertical axis shows the ratio with respect to the NANOGrav measurement. The discrepancy between the predicted and the three measured values ranges from 2--4.5$\sigma$. Red open symbols correspond to the scaling relations from MM13, while blue open symbols correspond to KH13. The open circles, crosses, and diamonds use the VDF from B10, B11, and T20, respectively, while the open stars were computed using the GSMF from DS15.}
    \label{fig:hc_comparison}
\end{figure}

\section{Can direct SMBH observations and GWs be reconciled?}\label{sec:discussion}
The predicted characteristic strain at $f=$yr$^{-1}$ for all of the combinations of velocity dispersion and stellar mass functions with scaling relations described in Sec.~\ref{sec:scaling}, and the values measured by PTAs are shown in Fig.~\ref{fig:hc_comparison}. The discrepancy between the different predictions and measurements varies between $2-4.5\sigma$, assuming Gaussian errors. Based on the discussion presented in Sec.\ref{sec:PTA}, we explore two aspects of the SGWB prediction that may be subject to change: the SMBH mass function and the merger history. We find that an increase in the merger rate of black holes is insufficient to boost the predicted characteristic strain to the value measured by PTAs. 

Motivated by the general agreement between the black hole mass density inferred from local scaling relations and from the quasar luminosity function through So\l{}tan's argument for a radiative efficienty around $\epsilon_r\sim 0.1$, we begin by considering a scenario in which we require the local mass density of BHs to be unchanged. We also consider more general variations to the BH abundance, such as the existence of a new population of SMBHs, and infer their required mass in order to match the PTA measurement. 

\begin{figure}
    \centering
    \includegraphics[width=0.42\textwidth]{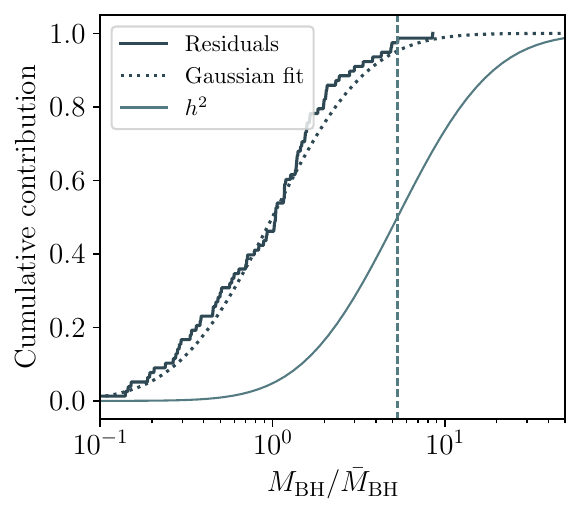}
    \caption{Cumulative contribution of residuals of the $M-\sigma$ relation. The dark blue solid line shows the residuals of the scaling relation fit from the data presented in Ref.~\cite{mcconnell_ma}, the dotted line shows a Gaussian fit to the residuals, and the light blue solid line shows the cumulative contribution to the characteristic strain from the same fit. Note that the scatter in the observed data includes both an intrinsic component and a contribution from measurement errors. The results shown above include both and therefore correspond to a slightly larger scatter (0.43) than the intrinsic (0.38).}
    \label{fig:residual}
\end{figure}

\begin{figure*}[t]
    \centering
    \includegraphics[width=0.85\textwidth]{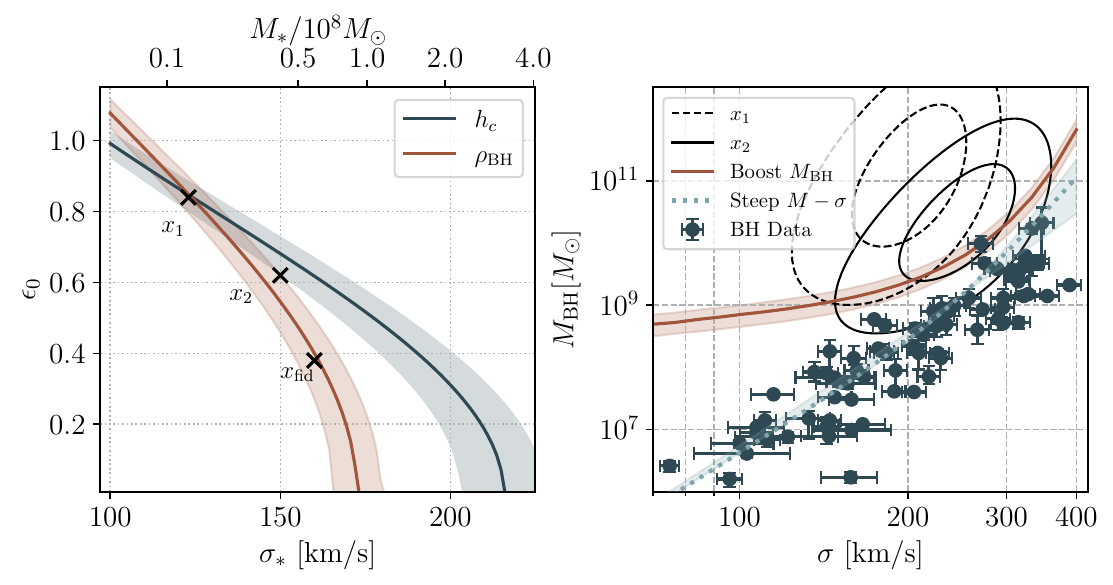}
    \caption{Parameter values required to match both the measured characteristic strain from PTAs and the inferred black hole mass density. The panel on the left shows the curves of constant $h_c$ (in dark blue) and $\rho_{\rm BH}$ (in red). The best-fit parameters from direct observations of SMBHs and host-galaxy properties correspond to the point $x_{\rm fid}=(159.6, 0.38)$. The points $x_1=(123, 0.84)$ and $x_2=(150, 0.62)$ correspond to sets of parameters where the predicted $h^2_c$ and $\rho_{\rm BH}$ exactly match the central fiducial value, and where they match within a 90\% confidence interval, respectively. The points in the panel on the right are the BH data used in Ref.~\cite{mcconnell_ma}. The inner (outer) contours show the regions where 50\% (90\%) of the contribution to the SGWB come from for the set of parameters $x_1$ in dashed and $x_2$ in solid lines. The red curve corresponds to the fiducial VDF and $M-\sigma$ relation, but with an additional population of BHs associated with galaxies of a given $\sigma$. Each point along the red curve therefore shows the required BH mass if a single BH population were to account for the $h_c$ value measured by NANOGrav. The light blue dotted line and shaded band shows Eq.~\ref{eq:double_pl} for $d_{\bullet}=10.5^{+1.5}_{-3.6}$.}
    \label{fig:match_hc_rho}
\end{figure*}

\subsection{Changes to galaxy scaling relations}

As noted in Sec.~\ref{sec:PTA}, if we require the characteristic strain to be fixed to its measured value and the BH mass density to be consistent with the current estimates, then the relevant outcome of any variation of the VDF or the $M-\sigma$ parameters is to change the mass that most contributes to the SGWB. Fig.\ref{fig:hc_rho_ratio} therefore allows us to infer that the characteristic strain must be dominated by $\sim 3\times 10^{10} M_{\odot}$ black holes, roughly 10 times larger than previously predicted. 

We explore variations in parameters that lead to such enhancements in the abundance or the typical masses of the most massive SMBHs, such as the intrinsic scatter $\epsilon_0$ in the $M-\sigma$ relation and the characteristic velocity dispersion $\sigma_*$. A larger $\sigma_*$ boosts the abundance of high velocity dispersion galaxies, while a higher intrinsic scatter in the $M-\sigma$ relation allows for more massive black holes to be associated with galaxies of a given $\sigma$. It is worth noting that, since the characteristic strain is weighted by $M^{5/3}$, it will always receive contributions from more massive outliers of the $M-\sigma$ relation, even for the best-fit value of $\epsilon_0$ from local observations. This feature is shown in Fig.~\ref{fig:residual}, using the black hole data from Ref.~\cite{mcconnell_ma}. Assuming the best-fit value of scatter results in 50\% of the contribution to $h_c$ being sourced from black holes that lie in the 5\% tail of the $M-\sigma$ relation.

We show in Fig.~\ref{fig:match_hc_rho} the curves of constant black hole mass density and characteristic strain in the $\sigma_*$ and $\epsilon_0$ parameter space. Requiring both values to be consistent with current measurements therefore requires particular values of such parameters, which correspond to the intersection of the curves shown. We highlight two points of intersection: $x_1$, which results in an exact match to the central value of both the measured $h^2_c$ and inferred $\rho_{\rm BH}$, while $x_2$ corresponds to a set of parameters that lead to predicted values of $h^2_c$ and $\rho_{\rm BH}$ that are consistent with observations at a 90\% confidence level.

The point $x_2$ has a very similar value of $\sigma_*$ to the fiducial, but a significantly larger intrinsic scatter in the $M-\sigma$ relation. Since the effect of scatter on the final SMBH mass function is to increase the abundance of the most massive black holes, this appears to be a straightforward mechanism to boost the characteristic strain, while minimizing the change to the BH mass density. However, the required value of scatter is in significant disagreement with local observations. This can be seen in the residuals of the $M-\sigma$ relation and, equivalently, in the right panel of Fig.~\ref{fig:match_hc_rho}. The points with error bars show the data set used in Ref.~\cite{mcconnell_ma}, while the contours show the regions that contribute 50\% and 90\% of the SGWB. This illustrates that changing the parameters of the VDF or the $M-\sigma$ relation in order to increase the GW prediction leads to the conclusion that the majority of the contribution to the SGWB would be sourced by host galaxies with associated SMBHs that have never been observed.

Motivated by the apparent upwards trend of the $M-\sigma$ relation for high velocity dispersion galaxies, we implement an $M-\sigma$ relation given by the double power law
\begin{equation}
    M_{\rm BH} = 10^{a_{\bullet}}\left(\frac{\sigma}{200 \text{km s}^{-1}}\right)^{b_{\bullet}} + 10^{c_{\bullet}}\left(\frac{\sigma}{200 \text{km s}^{-1}}\right)^{d_{\bullet}},
    \label{eq:double_pl}
\end{equation}
where $a_{\bullet}$, $b_{\bullet}$, and a single value of scatter $\epsilon_0$ are fixed to the values fit in MM13, and $c_{\bullet}$ is fixed such that the two power laws intercept at $\sigma=250$km s$^{-1}$. In order to match $h_c(f=\text{yr}^{-1})=2.4^{+0.7}_{-0.6}\times 10^{-15}$, we find $d_{\bullet}=10.5^{+1.5}_{-3.6}$. It has been argued in the literature that an upwards trend in the $M-\sigma$ relation may be driven by the velocity dispersion saturating at high values (which is not expected or observed for the other host galaxy properties), and therefore that bulge mass or luminosity scaling relations offer a more reliable prediction of the massive end of the black hole mass function. However, it is worth emphasizing that since we find a consistent value of the characteristic strain for predictions based on $\sigma$ and $M_{\rm bulge}$, then the steepening required to match the PTA measurements must necessarily be larger than this effect, as it would also require a boost in the $M_{\rm bulge}$-based prediction.


Although the BH mass density has been a useful point of comparison between SMBH mass functions estimated from local scaling relations and inferences based on the quasar luminosity function, it is not a directly measured quantity. If we instead allow $\rho_{\rm BH}$ to assume any value, $\sigma_*$ and $\epsilon_0$ may lie in any point along the dark blue curve in Fig.~\ref{fig:match_hc_rho}. We may also suppose that the measured VDF and $M-\sigma$ relation are fixed, but that there is a new population of SMBHs with a typical mass $M_{\rm new}$ hosted by galaxies with $\sigma_{\rm new}$, in addition to the standard population. The red curve on the right panel in Fig.~\ref{fig:match_hc_rho} shows the required value of $M_{\rm new}$ as a function of $\sigma=\sigma_{\rm new}$\footnote{We assume a Gaussian bump centered at each value of $\sigma$ with width of 10km s$^{-1}$ on top of the standard $M-\sigma$ relation, boosting SMBH masses by the amount required to match $h_c$ measured by NANOGrav. The red curve shows the required SMBH mass when galaxies with each value of $\sigma$ are boosted.}. As expected from our previous discussion, boosting the masses of SMBHs near the peak of the mass kernel of the SGWB leads to the smallest change relative to its fiducial value, while high velocity dispersion hosts require SMBHs of increasing masses to overcome their decreasing abundances and the required SMBH masses associated with low velocity dispersion hosts plateaus at $\sim 10^9 M_{\odot}$ as a consequence of the plateau in the VDF, leading to values that are in significant disagreement with observations.

In Appendix \ref{app:local} we compare the population of SMBHs required to explain the PTA results with those measured locally and inferred using local scaling relations. Fig.~\ref{fig:Nbh} (also Fig. \ref{fig:BHMF}) shows that the best-fit PTA results can be explained by a SMBH space-density in the local universe corresponding to a yet-undetected population of order 10 $3 \times 10^{10} M_\odot$ SMBHs within 100 Mpc (a volume inspired by the MASSIVE survey of massive galaxies and SMBHs \cite{2014ApJ...795..158M}).  This required space-density of very massive BHs is independent of whether the population is due to non-Gaussian scatter in BH-galaxy correlations or the steepening in such correlations highlighted above.

\subsection{Changes to the merger history}\label{sec:merger_history}

\begin{figure*}[t]
    \centering
    \includegraphics[width=0.85\textwidth]{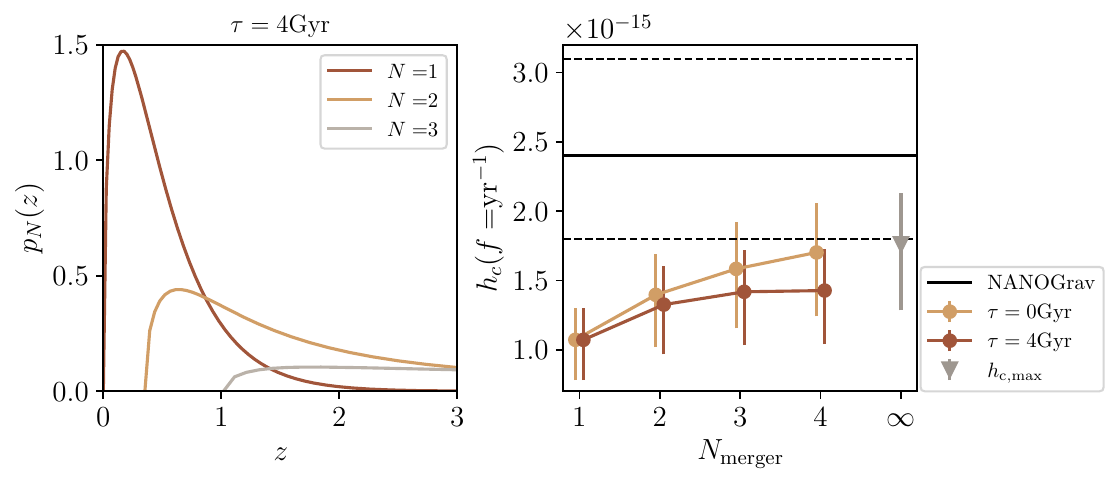}
    \caption{Impact of multiple merger events on the predicted characteristic strain. The panel on the left shows the redshift distribution of the first 3 merger events, assuming a fixed timescale of $\tau=4$ Gyr between subsequent mergers. The panel on the right shows the cumulative increase of the predicted characteristic strain as a function of number of mergers. The orange points show the increase for instantaneous mergers, the red points correspond to a 4 Gyr time-delay, and the grey arrow shows the upper limit computed in Eq.~\ref{eq:eqmass_boost}.}
    \label{fig:Nmergers}
\end{figure*}

In the previous section, we assumed that all black holes in the present-day Universe have undergone one merger event in order to connect the single SMBH mass function to the binary total mass function. Here we investigate whether changes to the merger history of SMBHs can sufficiently increase the amplitude of the SGWB.

We begin with simple and unphysical, but informative example. Ignoring the time scale for the black holes to merge and assuming that all mergers occur between equal mass black holes, we can estimate the boost to the characteristic strain from multiple mergers by noting that at each earlier generation $N$, the number of black holes will double and their masses will be halved. That is, we assume that the present-day single black holes of masses $M$ were formed from from equal-mass parents of masses $M/2$ that were twice as abundant, and so on. This leads to the following convergent series
\begin{equation}
    \left. h^2_{c}\right|_{N\rightarrow \infty} = \left. h^2_{c}\right|_{N=1} \left(\sum_{N=1}^{\infty} 2^{-2(N-1)/3}\right) \approx 2.7 \left.h^2_{c}\right|_{N=1}.\label{eq:eqmass_boost}
\end{equation}
Hence, even in the extreme scenario of an infinite number of instantaneous mergers, the characteristic strain $h_c$ can only be a factor of 1.64 larger than our previously predicted value, which is still lower than the best-fit value of any of the published PTA results.

Instead of an upper limit, we will now compute a fiducial estimate of this boost, assuming that they occur with a certain distribution of mass ratios and redshifts. We consider two compounding effects at each previous merger: the shift of the redshift distribution and the reduced efficiency of GW emission relative to an equal mass merger. We assume that the redshift distribution for the latest merger $p_{N=1}$ is given by Eq.~\ref{eq:pz_pq}, and compute the distribution of the previous merger $p_{N+1}$ by considering a fixed timescale $\tau$ between subsequent mergers. Every redshift $z$ can therefore be mapped onto the redshift of the previous merger $z_{N+1} = z_{N}+\Delta z(\tau)$. Since some fraction of objects will never merge (i.e. $t(z) + \tau$ may be larger than the age of the Universe), the new distribution function is normalized to the fraction that merges, which is not necessarily 1. We show in Fig.~\ref{fig:Nmergers} a case with $\tau=4$ Gyr and the resulting redshift distribution of each previous merger.

The effect of unequal mass mergers can be computed by considering that the progenitors of each black hole with a final mass $M_{\rm BH}$ will have masses $M_{\rm BH}/(1+q)$ and $qM_{\rm BH}/(1+q)$, where $q$ is drawn from the distribution function defined in Eq.~\ref{eq:pz_pq}. We show in App.~\ref{app:eff_q} that the boost to the characteristic strain after $N$ mergers predicted through a Monte Carlo sampling of the mass ratio distribution is nearly identical to the prediction with a fixed mass ratio given by the mean $\avg{q}$. Furthermore, we derive an equivalent series to Eq.~\ref{eq:eqmass_boost} for the unequal mass case. The highest boost is obtained for $q=1$ given by \ref{eq:eqmass_boost} but it does not change much with $q$, in the limit $q\rightarrow 0$ the boost factor is $12/5=2.4$.

The combined effect of the finite merger timescale and the unequal mass mergers is a smaller boost to the characteristic strain as a function of number of mergers and a convergence to a lower maximum value, since the finite merger time implies a cutoff to the contributions. The cumulative contribution of mergers is shown in the right panel of Fig.~\ref{fig:Nmergers}. The timescale of 4 Gyr is inspired by the state-of-the-art simulations presented in Ref.~\cite{Kelley:2016gse}, which is the average lifetime of a $\sim 10^9 M_{\odot}$ black hole. Note that this is still optimistic in several ways. For instance, it neglects the timescale of the galaxy/halo merger and that black holes grow through accretion, which would result in a significantly lower mass for the parent SMBHs at earlier times. We therefore conclude that, while an increase in the merger rate may alleviate the tension between the predicted amplitude of the SGWB and the values measured by PTAs, it is insufficient to fully account for the discrepancy. 

\section{Conclusion}\label{sec:conclusion}
The detection of nanohertz gravitational waves by several PTA collaborations has recently opened a new window into the supermassive black hole population, that is complementary to the picture offered by existing direct local measurements and by quasars in the distant Universe. The GW measurement is particularly sensitive to the most massive SMBHs, which are rare and therefore challenging to find through direct observations. 

By comparing GW and electromagnetic measurements of SMBHs (following \cite{Phinney:2001di}), we find an inconsistency in the inferred SMBH population, leading to a lower predicted value of the SGWB for mass functions consistent with local observations. We show that this difference cannot be accounted for by boosting the merger rate of SMBHs and is therefore likely a consequence of an underestimated mass function of SMBHs. We show that if the current estimate of the black hole mass density, consistent with the integrated luminosity of quasars through the So\l{}tan argument, is to be preserved, then the PTA measurement generally implies that the SMBH mass that most contributes to the SGWB must be $\sim 3 \times 10^{10} M_\odot$, about ten times higher than would be naively predicted using local measurements of the SMBH mass function. This would suggest that the most massive SMBHs found in our local Universe comprise the typical population contributing to the SGWB.   

We explore a few potential avenues that could lead to a higher prediction of the high mass tail of the SMBH mass function. We begin by modifying the intrinsic scatter in the $M-\sigma$ relation and the turnover scale in the VDF, but find that the required parameters to account for the PTA measurement are inconsistent with the residuals of the $M-\sigma$ fit. We then show that a steeper $M-\sigma$ relation for high velocity dispersion galaxies can account for the missing high-mass population. Although it has been argued that the $M-\sigma$ relation should steepen as a consequence of the saturation of the velocity dispersion at high values in gas-poor mergers, we emphasize that since both the $\sigma$- and $M_{\rm bulge}$-based predictions lead to consistent and underpredicted values of the characteristic strain, then the required boost to the abundance of the most massive black holes must necessarily be larger than this effect. This exploration leads to a directly testable prediction for the local abundance of the most massive SMBHs and therefore highlights the importance of populating this region of parameter space with local observations. Quantitatively, the best-fit PTA results require a SMBH space-density in the local universe that corresponds to a yet-undetected population of order 10 $3 \times 10^{10} M_\odot$ or 1 $10^{11} M_\odot$ SMBHs within 100 Mpc (a volume in which stellar and gas dynamical BH measurements are feasible).  However, masses significantly above a few $10^{10} M_\odot$ are disfavored since they would likely be detected as individual sources by PTAs.

Although the GW signal alone offers only an integrated measurement of the SMBH mass function, we showed that combining GWs with local observations of SMBHs and the quasar luminosity function leads to a greater insight about this black hole population. Crucially, GW measurements probe the massive end of the SMBH mass function, which are rare and therefore elusive in electromagnetic observations. While some black holes of masses $\sim 10^{10}M_{\odot}$ have been found locally, the low surface brightness of the cored inner region hinders direct mass measurements. Furthermore, since these extremely massive SMBHs are most likely hosted by galaxies that appear to have been formed from repeated gas-poor mergers, they are also unlikely to be seen as the brightest quasars. GWs may therefore be offering a unique perspective on the most massive black holes in the Universe.

\acknowledgments
We would like to thank Jenny Greene, Scott Tremaine, Luke Zoltan Kelley, and Nick Kokron, for helpful discussions. EQ and MZ thank Roger Blandford for a cryptic comment about the NANOGrav results at the Simons Symposium on Multiscale Physics, which motivated this work. GSP gratefully acknowledges support from the Friends of the Institute for Advanced Study Fund. MZ is supported by NSF 2209991 and NSF-BSF 2207583.  EQ is supported in part by a Simons Investigator Award from the Simons Foundation.

\bibliography{ref.bib}
\bibliographystyle{utcaps}

\appendix

\begin{figure}[h]
    \centering
    \includegraphics[width=0.45\textwidth]{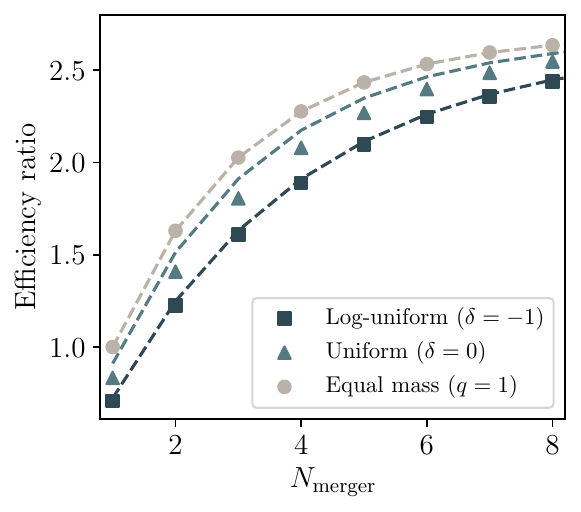}
    \caption{Efficiency of GW emission relative to the equal mass case. Symbols corresponds to the results of merger tree simulations and the lines the series of equation \ref{eq:series-fixedq}.}
    \vspace{-1cm}
    \label{fig:enter-label}
\end{figure}

\section{Efficiency of unequal mass mergers}\label{app:eff_q}
Assume all mergers have a fixed mass ratio $q$ and corresponding $\eta$. Let us define $S(N,q)$ the efficiency ratio relative to a single equal mass merger of a tree of mergers that goes down $n$ levels. All merger have an efficiency ratio relative to the equal mass case of $4 \eta$. After one merger the two progenitor black holes have masses $M_{\rm BH}/(1+q)$ and $M_{\rm BH}q/(1+q)$ so:
\begin{equation}
    S(0,q) = 4\eta \\\\ ; \\\\ S(1,q) = 4\eta \left(1+\frac{1}{(1+q)^{5/3}}+\frac{q^{5/3}}{(1+q)^{5/3}} \right).
\end{equation}
After a second level, the black holes split again and there is one black hole with mass $M_{\rm BH} 1/(1+q)^2$, two with masses $M_{\rm BH} q/(1+q)^2$, and one with $M_{\rm BH}(q/(1+q))^2$, so that:
\begin{equation}
    S(2,q) = 4\eta \left(1+\frac{1+q^{5/3}}{(1+q)^{5/3}} + (\frac{1+q^{5/3}}{(1+q)^{5/3}})^2 \right).
\end{equation}
To go down each one level, each of the black holes split into pieces by factors $1/(1+q)$ and $q/(1+q)$, so there are many possible masses. The number of black holes of the different masses however are given by the binomial coefficients so that in general:
\begin{equation}
    S(N,q) = 4\eta \sum_{n=0}^N\left(\frac{1+q^{5/3}}{(1+q)^{5/3}} \right)^n= 4\eta \frac{1-\left(\frac{1+q^{5/3}}{(1+q)^{5/3}} \right)^{N+1}}{1- \left(\frac{1+q^{5/3}}{(1+q)^{5/3}} \right)}
    \label{eq:series-fixedq}
\end{equation}
For a fixed $N$, $S(N,q)$ is maximum for $q=1$. 

Rather than label the tree by the number of levels, one could label it by the mass of the largest black hole in that level, $M_{\rm BH} /(1+q)^{N+1}$.  The analytic prediction in the figure simply plots $(1+q)^{N+1}$ vs $S(N,q)$ where $q$ is the average value for each distribution, where $\langle q \rangle = [0.3909,0.55,1]$ for the log, linear and equal mass cases (assumes $q_{min}=0.1$). In these formulas $N$ does not need to be an integer. The discrete points show the results of averages of Monte Carlo realization where the mass ratio was sampled form the appropriate distribution. Each subsequent point corresponds to one additional level of the tree.

\section{Implication for local observations}
\label{app:local}
\begin{figure}[t]
    \centering
    \includegraphics[width=0.45\textwidth]{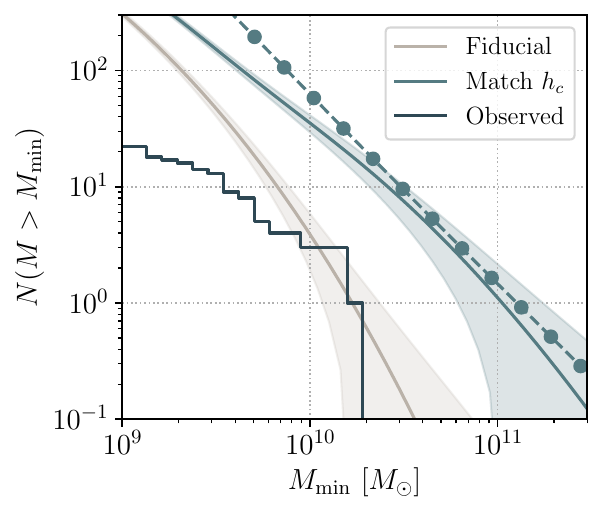}
    \caption{Total number of SMBHs above a given minimum mass $M_{\rm min}$. Solid lines with error bands correspond to predictions using the fiducial VDF and $M-\sigma$ relation (grey) and the fiducial VDF and an $M-\sigma$ relation given by Eq.~\ref{eq:double_pl} with the best-fit value that matches the characteristic strain measured by PTAs (light blue). The dashed line with dots correspond to the fiducial mass function, with a Gaussian bump above $M_{\rm min}$ such that the amplitude is chosen to match the measured $h_c$. The dark blue curve shows the SMBHs observed so far.}
    \label{fig:Nbh}
\end{figure}

The higher amplitude of the SGWB found by PTA observations may imply a ``missing" populations of SMBHs in the current estimates of the mass function. While the PTA signal alone cannot distinguish their properties, requiring consistency with the current estimate of the black hole mass density would suggest an increase in the high-mass end of the SMBH mass function. If this is the case, we should expect an increase in the number of very massive SMBHs in nearby galaxies and we provide a rough estimate of this population. 

Inspired by the ongoing observations of the MASSIVE Survey~\cite{2014ApJ...795..158M}, we show in Fig.~\ref{fig:Nbh} the number of SMBHs above a certain minimum mass $M_{\rm min}$ within a survey with a spherical volume of 100 Mpc radius. The solid lines correspond to predictions for a single mass function, the lower computed with out fiducial VDF and $M-\sigma$ relation, and the higher with the fiducial VDF and a steeper $M-\sigma$ relation that reproduced the central value of $h_c$ observed by NANOGrav, and error bands which correspond to Poisson errors. The dashed line corresponds to a SMBH mass function identical to the fiducial, but with a Gaussian bump added above $M_{\rm min}$ with width of 0.1 dex and with a height chosen so as to match $h_c$. Each point in the dashed line therefore corresponds to a different mass function, and should be interpreted as a statement that the characteristic strain observed by NANOGrav may be explained by a mass function that predicts $\sim 60$ SMBHs with masses above $\gtrsim 10^{10} M_{\odot}$ or $\sim 2$ SMBHs with masses above $\gtrsim 10^{11} M_{\odot}$ within a distance of 100Mpc. However, masses significantly above $10^{10} M_{\odot}$ are disfavored by upper limits in individual SMBHB searches, as they would likely be detected as point sources in PTA observations \cite{NANOGrav:2023pdq}. The line of observed black holes includes the data from Ref.~\cite{mcconnell_ma}, with the addition of the SMBH reported in Ref.~\cite{2016Natur.532..340T}. Note that this does not correspond to a complete sample within the assumed volume, unlike the sample measured by the MASSIVE Survey.

\end{document}